\documentclass[twocolumn,showpacs,preprintnumbers,groupedaddress,amsmath,amssymb,prb,superscriptaddress]{revtex4}

\usepackage{graphicx}
\usepackage{color}

\newcommand{\expct}[2]{\left\langle #1 \right\rangle_{#2}}
\newcommand{\expcts}[2]{\langle #1 \rangle_{#2}}

\newcommand{\cumus}[2]{\langle\!\langle #1 \rangle\!\rangle_{#2}}
\newcommand{\abs}[1]{\left\vert #1\right\vert}
\newcommand{\C}{\mathcal{C}}
\newcommand{\TC}{T_{\C}\, }
\newcommand{\T}{\mathcal{T}}
\newcommand{\Exp}[1]{\text{e}^{#1}}
\newcommand{\I}{\text{i}}
\newcommand{\D}{\text{d}}
\newcommand{\ie}{{\it i.e.}}
\newcommand{\eg}{{\it e.g.}}

\begin{document}

\title{Charge transfer statistics of a molecular quantum dot\\
with strong electron-phonon interaction}
\author{S.~Maier}
\affiliation{Institut f\"ur Theoretische Physik,
Ruprecht-Karls-Universit\"at Heidelberg,\\
 Philosophenweg 19, D-69120 Heidelberg, Germany}
 \author{T.~L.~Schmidt}
\affiliation{Department of Physics, Yale University, 217 Prospect Street, New Haven, Connecticut 06520, USA}
\author{A.~Komnik}
\affiliation{Institut f\"ur Theoretische Physik,
Ruprecht-Karls-Universit\"at Heidelberg,\\
 Philosophenweg 19, D-69120 Heidelberg, Germany}
\date{\today}

\begin{abstract}
We analyze the nonequilibrium transport properties of a quantum dot with a harmonic degree of freedom (Holstein phonon) coupled to  metallic leads, and derive its full counting statistics (FCS). Using the Lang-Firsov (polaron) transformation, we construct a diagrammatic scheme to calculate the cumulant generating function. The electron-phonon interaction is taken into account exactly, and the employed approximation represents a summation of a diagram subset with respect to the tunneling amplitude. By comparison to Monte Carlo data the formalism is shown to capture the basic properties of the strong coupling regime.
\end{abstract}


\maketitle

In the past decades, the miniaturization of electric circuits has crossed the divide between the microscale and the nanoscale. Molecular and atomic electronics are no longer mere theoretical concepts. A wide variety of experimental setups has been developed for the exploration of the electronic and mechanical properties of nanometer-sized objects like carbon nanotubes, $C_{60}$-fullerenes and complex molecules. It has become possible to connect these samples to mesoscopic environments and to investigate their transport properties.\cite{ParkMcEuen,smit2002,Zhitenev2002,ParkRalph,Qiu2004,Yu2004,YouNatelson,Pasupathy2005,djukic05,Sapmaz2006, Leturcq2009}

The electronic structure of these nanosized objects is best captured by the concept of a quantum dot, \ie, by an arrangement of energy levels which correspond to the molecular orbitals of the device. One of the most prominent and fundamental models for the theoretical description of quantum dots is the Anderson impurity model, which accounts for the tunnel coupling to noninteracting electron reservoirs and for the local Coulomb interaction between the electrons populating the quantum dot.\cite{Anderson1961} In the case of contacted molecules, where charging is often accompanied by structural deformations of the molecule itself, however, this model is often an oversimplification. For a more realistic description, an explicit consideration of the coupling to vibrational degrees of freedom is necessary. This is accomplished by the Anderson-Holstein model (AHM).\cite{Holstein1959,Hewson2002}

In its full extent, the AHM captures a huge variety of physical phenomena. Its physical properties depend on several energy scales, \eg, temperature, charging energy, hybridization energy, level spacing and electron-phonon interaction strength. These define many interesting and physically distinct regimes in parameter space. In this paper we are mainly interested in the effect of electron-phonon interactions on the charge transport through a contacted molecule. The model can therefore be simplified to contain a single electronic level (thus neglecting the spin degree of freedom as well as the charging energy) linearly coupled to a local (Holstein) phonon, \ie, a bosonic oscillator degree of freedom with a single frequency. Even this simplified model, which in the following will be referred to as AHM, offers rich physics. Whereas the conductance and the nonlinear $I-V$-characteristic of such a system can be approached by a number of methods, such as
diagrammatic Monte Carlo schemes,\cite{Muhlbacher2008} rate equations,\cite{koch05,Leturcq2009} perturbation theory,\cite{Vega2006,galperin06,Flensberg2003,Riwar2009,Tahir2010} and $P(E)$ theory,\cite{kast2010} its full counting statistics (FCS) is well understood only in the limit of weak electron-phonon coupling.\cite{Vega2006,Paulsson2005,Schmidt2009,Egger2008,Haupt2009,Avriller2009,Urban2010}

\begin{figure}[ht]
\includegraphics{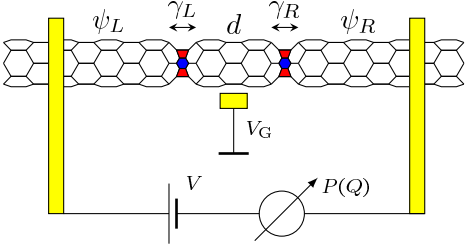}
\caption{(Color online) Sketch of a suspended carbon nanotube quantum dot. The fields $\psi_{L,R}(x)$ describe semi-infinite leads, and $d$ is the electron annihilation operator on the quantum dot. The gate voltage $V_G$ allows for an adjustment of the energy levels of the dot. The bias voltage $V$ induces a finite charge current across the dot. $P\left(Q\right)$ denotes a measurement apparatus for the charge transfer statistics.}
\label{fig:setup}
\end{figure}

In this paper we would like to extend these results and present a calculation of the FCS beyond the weak-coupling limit. One possible experimental setup in which strong electron-phonon interaction can be reached is a quantum dot embedded in a suspended carbon nanotube,\cite{PhysRevLett.96.026801,BenjaminLassagne08282009,Leturcq2009} as depicted in Fig.~\ref{fig:setup}. The electrodes as well as the quantum dot are made of a single carbon nanotube subject to a bias voltage $V$. The electronic level structure of the quantum dot can be tuned by an additional backgate. Because of its simple structure, the vibrational modes of such a quantum dot are well understood.\cite{mariani2009} However, the model employed is fairly general. Depending on the parameter regime, it also allows the desription of transport through molecules contacted using mechanically-controlled break junctions\cite{smit2002,djukic05} and STM tips,\cite{Qiu2004,stipe98} as well as in nanoelectromechanical setups.\cite{knobel03}

The starting point of our calculations is the Hamiltonian of the AHM,
\begin{gather}          \label{H0}
H=H_0 + H_\text{T} + H_\text{el-ph}\text{.}
\end{gather}
$H_0$ is the Hamiltonian for the uncoupled degrees of freedom of the electrodes, the single-level quantum dot, and the bosonic mode,
\begin{eqnarray}
H_0 &=
H_L[\psi_L] + H_R[\psi_R] + \epsilon_0 d^\dagger d + \Omega b^\dagger b\text{.}
\end{eqnarray}
The first two terms describes the electrodes in the language of noninteracting electron field operators $\psi_{L, R}\left(x\right)$ which are held at chemical potentials $\mu_{L,R}$. This is achieved by a bias voltage $V=\mu_L - \mu_R$ applied across the contacts (we use natural units $e=\hbar=k_\text{B} =1$ throughout). In the case of noninteracting leads, the only property of $H_{L,R}$ which is necessary for further calculations is the local tunneling density of states $\rho(\omega)$. As a simplification, we work in the wide flat band limit $\rho(\omega)=\rho_0$, which corresponds to free fermions $\psi_{L,R}$ with linear dispersion and infinite bandwidth, but any other shape of $\rho(\omega)$ can be treated in the same way. The third term in $H_0$ comes from the single electronic level of the quantum dot with bare energy $\epsilon_0$ and associated creation and annihilation operators $d^\dag$ and $d$. The last term in $H_0$ describes a phonon mode with frequency $\Omega$ and bosonic creation and annihilation operators $b^\dag$ and $b$.

The tunneling Hamiltonian $H_{\text{T}}$ describes the hopping of electrons from the leads to the dot and back:
\begin{gather}
H_{\text{T}}= \frac{\gamma}{2} \sum_{m=L,R}\left[  \psi^\dagger_m\left(x=0\right) d + \text{h.c.}\right]\text{,}
\end{gather}
where $\gamma$ is the tunneling amplitude and we assumed symmetric coupling to both leads. An asymmetric coupling requires only small adjustments in the calculations but significantly complicates notation. The last term in Eq.~(\ref{H0}) is the electron-phonon interaction,
\begin{gather}
H_\text{e-ph}=g d^\dagger d\left(b^\dagger + b\right)\text{.}
\end{gather}
It linearly couples the dot occupation operator $d^\dag d$ to the displacement of the harmonic oscillator $Q\sim b^\dag + b$ with a coupling strength $g$.

The fundamental quantity describing low-frequency electronic transport is the probability distribution function $P\left(Q\right)$ of transferring $Q$ units of charge during a measurement time $\mathcal T$, which we assume to be the longest time scale. Physical observables can then be calculated as averages with respect to this distribution function. The expectation value $\expct{Q}{}$ is related to the average current, $\expcts{Q}{}= \expct{I}{}\T$. The second cumulant $\cumus{Q^2}{} = \expcts{Q^2}{} - \expcts{Q}{}^2$ is directly related to the noise power $\cumus{Q^2}{}=S\T$, where $S = \frac{1}{2} \int d\omega \cumus{ I(t) I(0) + I(0) I(t) }{}$. Instead of calculating the probability distribution function itself it is often more convenient to calculate its cumulant generating function (CGF) $\ln\chi\left(\lambda\right) = \ln\sum_Q \Exp{\I\lambda Q} P\left(Q\right)$. It has been shown,\cite{levitov1996,nazarovlong} that $\ln \chi(\lambda)$ can be expressed in terms of Keldysh Green's functions (GFs) and this formalism has been successfully applied to a wide range of transport problems. The fundamental expression for calculating the CGF is\cite{levitov04}
\begin{gather}
  \chi\left(\lambda\right) = \expct{ \TC\, \Exp{-\I\int_\C \D t\, T_\lambda\left(t\right)}}{0} \, ,
\end{gather}
where $\TC$ denotes time-ordering along the Keldysh contour $\C$ and the expectation value is written in the interaction picture with respect to the Hamiltonian $H_0 + H_\text{el-ph}$. The tunneling operator $T_\lambda$ is given by
\begin{gather}
T_\lambda = \frac{\gamma}{2} \left[ \Exp{\I\lambda/4} \psi^\dagger_L(0) d + \Exp{-\I\lambda/4} \psi^\dagger_R(0) d + \text{h.c.} \right].
\end{gather}
The counting field $\lambda$ is explicitly time-dependent on the Keldysh contour: $\lambda(t \in \C_\pm) = \lambda^\pm$. At the end of the calculation one has to replace $\lambda^+ = -\lambda^- =\lambda$. As was shown in Ref.~[\onlinecite{AndersonFCS}], in the limit $\T \to \infty$ the following expression holds
\begin{gather}
\frac{\partial}{\partial\lambda^-} \ln\chi\left(\lambda^-,\lambda^+\right) = -\I\T \expct{ \frac{\partial T_\lambda}{\partial \lambda^-}}{\lambda} \, .
\label{eq:adiapot}
\end{gather}
The $\lambda$-dependent expectation value is defined as
\begin{gather}
\left\langle \cdots \right \rangle_\lambda=\frac{\expct{ \TC\,\cdots \,\Exp{-\I\int_\C \D t\, T_\lambda(t)}}{0}}{\chi\left(\lambda^+,\lambda^-\right)}\, .
\end{gather}
We proceed by defining the \emph{exact} $\lambda$-dependent dot GF $D_\lambda$ and the local (taken at $x=0$) \emph{free} electrode GF $g_m$ (where $m = L,R$) in Keldysh space by
\begin{align}
 D_\lambda(t,t') &= -\I\expct{ \TC d(t)d^\dag(t')}{\lambda}, \notag \\
 g_m(t,t') &= -\I\expct{ \TC \psi_m(0,t) \psi^\dag_m(0,t')}{0},
\end{align}
Then, Eq.~(\ref{eq:adiapot}) can be expressed as the following convolution,
\begin{align}
& \frac{\partial\ln\chi(\lambda^-,\lambda^+)}{\partial\lambda^-}  =- \frac{\I\T\gamma^2}{4} \times\\
& \int\frac{\D\omega}{2\pi}\sum_{m}\biggl[\Exp{-\I\lambda}D^{-+}_\lambda(\omega)
 g_m^{+-}(\omega)
 -\Exp{\I\lambda}g_m^{-+}
 (\omega)D^{+-}_\lambda(\omega)\biggr]\notag \, \text{,}
\end{align}
where $\lambda=(\lambda^- -\lambda^+)/4$. The usual route for the limit of weak electron-phonon coupling would be an expansion of the dot GF in the interaction strength $g$. Alternatively, the strong coupling regime can conveniently be approached using a Lang-Firsov (polaron) transformation,\cite{LangFirsov} $U=\exp\left[ \alpha d^\dag d(b^\dag-b)\right]$. One can check that the Hamiltonian transforms to $U (H_0 + T_\lambda + H_\text{el-ph}) U^\dag = H_0' + T_\lambda' + H_\text{el-ph}'$, where
\begin{align}
H_0'&= H_{L} + H_R + \left(\epsilon_0+\alpha^2\Omega\right) d^\dagger d \notag \\
&+ \Omega b^\dagger b-\alpha\Omega d^\dag d\left(b+b^\dag\right), \notag \\
T_\lambda'&=\frac{\gamma}{2} \sum_{m=L,R=-,+}\left[ \Exp{-\I m \lambda/4} \psi^\dagger_m(0) \Exp{\alpha(b^\dag-b)} d + \text{h.c.}\right],\notag \\
H_\text{el-ph}'&= g d^\dag d\left(b+b^\dag\right) - 2\alpha g d^\dag d \,.
\end{align}
Setting $\alpha=g/\Omega$ removes the electron-phonon interaction and leads to the polaron shift of the dot energy, $\epsilon_0\to \epsilon_0 - g^2/\Omega$. The polaron shift can be taken care of by appropriate gating and, for ease of notation, we use $\epsilon_0$ to denote the shifted dot level from now on. Moreover, the transformed tunneling operator $T_\lambda'$ now contains the dressed dot operator
\begin{align}
 D := \Exp{g (b^\dag-b) /\Omega} d =: X d.
\end{align}
The operator $D^\dag$ creates a particle on the dot which is dressed by a phonon cloud $X$. The next step is to express the dot GF in the language of the new transformed Hamiltonian.\cite{Flensberg2003,zazunov2007} One finds
\begin{align}
D_\lambda(t,t') &=-\I \expct{\TC D(t) D^\dag(t')}{\lambda} \notag \\
&= -\I \expct{ \TC d(t) d^\dag(t') X(t) X^{-1}(t')}{\lambda} ,
\end{align}
where the $\lambda$-expectation value now has to be calculated using the rotated Hamiltonians $H_0'$ and $T_\lambda'$.
\begin{figure}[t]
\includegraphics[width=8cm]{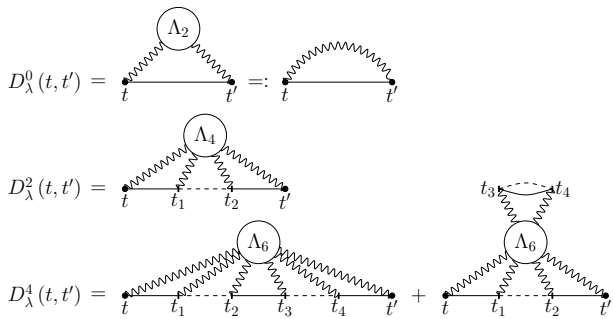}
\caption{Feynman diagrams for the contributions $D^0_\lambda\equiv D_0$, $D^2_\lambda$ and $D^4_\lambda$. The solid lines represent the free dot GF $d(t,t')$ and the dashed lines the free lead GF $g_{L,R}(t,t')$. The internal time variables $t_i$ are integrated over while $t,t'$ (denoted by the filled circles) are external time variables.}
\label{fig:perturb}
\end{figure}
In the next step we perform a formal expansion in the tunneling Hamiltonian $T_\lambda'$ and find
\begin{align}
D_\lambda\left(t,t'\right)&= \sum_{n=0}^\infty \gamma^{2n} D^{2n}_\lambda\left(t,t'\right) \, ,
\end{align}
where $D^{2n}_\lambda(t,t')$ is the sum of all terms to $2n$th order in the tunneling amplitude. As an example, the diagrammatic structure of the contributions $D^0_\lambda$, $D^2_\lambda$ and $D^4_\lambda$ is shown in Fig.~\ref{fig:perturb}. In addition to the free dot GF,
\begin{align}
 d(t,t') = -\I \expct{\TC d(t) d^\dag(t')}{0},
\end{align}
the calculation of the individual orders involves correlators of the form $\Lambda_{2n}(t_1,\ldots,t_{2n})=\expcts{ \TC\, X(t_1)X^{-1}(t_2)\ldots X(t_{2n-1}) X^{-1}(t_{2n})}{0}$. Since the $X$-operators are exponentials of free boson operators, the calculation of this correlation function is straightforward:
\begin{gather}
\Lambda_{2n}\left(t_1,\ldots,t_{2n}\right)=\prod_{i<j}^{2n}\Lambda\left(t_i-t_j\right)
\, ,
\end{gather}
where $\Lambda(t-t')$ is a function in Keldysh space,
\begin{gather}\begin{split}
\Lambda(t-t') &= \left\{\Lambda^{kl}(t-t')\right\}_{k,l=\pm}\\ &=\begin{pmatrix}
    \kappa(\abs{t-t'}) & \kappa(t' - t) \\
    \kappa(t-t')       & \kappa(-\abs{t-t'})
\end{pmatrix}\end{split} \, ,
\end{gather}
and $\kappa(t)$ is defined as
\begin{gather}
\kappa(t)=\exp{\left\{-\alpha^2\left[\left(\Exp{\I\Omega t} -1\right)n_\text{B} + \left(\Exp{-\I\Omega t}-1\right)\left(n_\text{B}+1\right)\right] \right\}}\text{.}
\end{gather}
The uncoupled phonon occupation number $n_\text{B} = \expcts{b^\dag b}{0}$ accounts for the initial occupation of the harmonic oscillator states. If the oscillator is coupled to a thermal environment,\cite{PhysRevB.69.245302} \eg, to a substrate or a backgate, $n_\text{B}$ is a temperature-dependent distribution function. In this case, it is often sensible to assume that it given by an equilibrium Bose distribution, $n_\text{B}(T)=(\Exp{\Omega/T}-1)^{-1}$, where $T$ is the bath temperature. In the limit $T\rightarrow 0$ $n_\text{B}$ approaches zero. Note that $n_\text{B}$ denotes the phonon number in the absence of coupling to the dot. It is generally different from the highly nontrivial, nonequilibrium occupation number that emerges as a consequence of the coupling to the dot, and which is governed by the transport processes in the lead-coupled system.\cite{PhysRevB.69.245302,Urban2010} The function $\kappa(t)$ can be expressed as a Fourier series
\begin{widetext}
\begin{gather}
\kappa\left(t\right) =\begin{cases}\Exp{-\alpha^2}\sum\limits_{n=0}^{\infty}\frac{\alpha^{2n}}{n!}\Exp{-\I n\Omega t} & T=0.\\
\Exp{-\alpha^2\left[2n_\text{B}(T)+1\right]}\sum\limits_{n=-\infty}^{\infty}I_n\left[2\alpha^2\sqrt{n_\text{B}\left(n_\text{B}+1\right)}\right]\Exp{n\Omega/2 T}\Exp{-\I n\Omega t}&T>0, \end{cases}
\end{gather}
\end{widetext}
where $I_n$ denotes the $n$th order modified Bessel function. There is a crucial difference between the $T=0$ and $T>0$ expansion. In the former case, only positive phonon numbers $n$ occur. This is natural because in this case, the phonons can only be excited. At finite $T$, in contrast, $n$ runs over positive and negative integers because now the phonons can be emitted or absorbed. In the limit $T\to 0$ both expressions have to coincide. This can be verified by considering the $x \to 0$ limit of the modified Bessel function $I_n(x) = x^n \{ [2^n \Gamma(n+1)]^{-1}+{\mathcal O}(x^2) \}$ where $\Gamma(n)$ denotes the gamma function.\footnote{$\Gamma(n+1)$ for negative integer $n$ is ill-defined, because the integral in the definition of the $\Gamma$-function does not converge. However, it is meaningful to formally set $\Gamma\left(n+1\right)=-\infty$ for integer $n<0$.}

\begin{figure}[t]
\includegraphics[width=8cm]{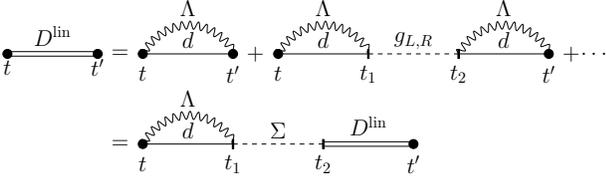}
\caption{Linear diagram resummation scheme. The single solid lines denote the free dot GF $d(t,t')$ and the dashed lines denote the free electrode GFs $g_{L,R}(t,t')$. Wiggly lines represent the phonon cloud propagator $\Lambda(t,t')$. This approximation allows to derive the Dyson equation in the second line, where $g_{L,R}$ acts as self-energy.}
\label{fig:subsum}
\end{figure}

\begin{figure}[t]
\includegraphics[width=8cm]{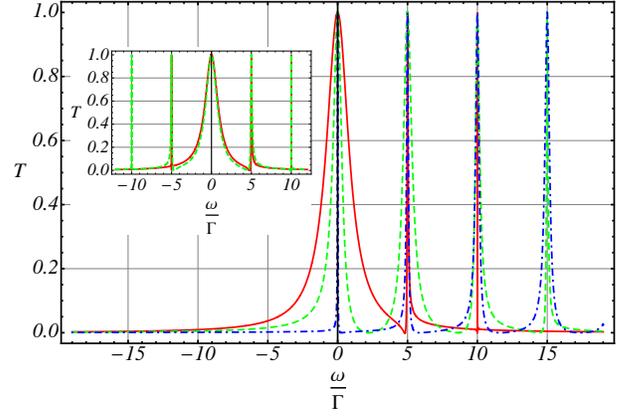}
\caption{(Color online) Transmission coefficient $T(\omega)$. \emph{Main graph:} Transmission coefficient for $T=\epsilon_0=0$, $\Omega/\Gamma=5$, and $g/\Gamma = 1,5,10$ (solid, dashed, dashed-dotted line, respectively). \emph{Inset:} Transmission coefficient for fixed $g/\Gamma=1$, $\Omega/\Gamma=5$ and $\epsilon_0=0$ for varying temperatures $T/\Gamma=1,10$ solid, dashed line.}
\label{fig:transcoeff}
\end{figure}

So far, no approximations have been performed. However, the complicated structure of the diagrams (see Fig.~\ref{fig:perturb}) makes an exact solution of the problem difficult. Moreover, a simple perturbative expansion in the tunneling amplitude $\gamma$ is inconvenient as the individual contributions are highly divergent. One of possible regularization methods is to couple the quantum dot to a fictitious bath to enforce a hybridization of the sharp dot level (aka Wigner-Weisskopf regularization).\cite{wigner1930} We take another path and calculate the dot GF using a linear approximation of the diagrammatic series (see Fig.~\ref{fig:subsum}). It allows us to derive the following Dyson equation in frequency domain,
\begin{gather}\begin{split}
\mathbf{D}^{\text{lin}}_\lambda &=\mathbf{D}_0 + \gamma^2 \mathbf{D}_0\left[\Exp{-\I\frac{\boldsymbol{\lambda}}{2}} \mathbf{g}_L +\Exp{\I\frac{\boldsymbol{\lambda}}{2}} \mathbf{g}_R \right]  \mathbf{D}_0+\ldots\\
     &=\mathbf{D}_0 + \mathbf{D}_0\boldsymbol{\Sigma}\mathbf{D}^{\text{lin}}_\lambda.\end{split}\label{eq:dyson}
\end{gather}
Here, we used the conventional matrix notation in Keldysh space, $[\mathbf{D}_0(\omega)]_{kl} = D_0^{kl}(\omega)$ for $k,l = \pm$, and similar for the self-energy $\boldsymbol{\Sigma}$. The function $D_0^{kl}(\omega)$ is the convolution of $\Lambda^{kl}(\omega)$ with the free dot propagator $d^{kl}(\omega)$, \ie, $D_0^{kl}(t,t') = \Lambda^{kl}(t,t') d^{kl}(t,t')$. In this approximation the dot GF can be determined exactly. A lengthy but straightforward calculation reveals that the CGF has the form of the Levitov-Lesovik formula\cite{levitov1996}
\begin{gather}\begin{split}
      \ln\chi(\lambda)&=\T \int\frac{\D\omega}{2\pi} \ln\biggl\{ 1+T(\omega) \\ &\times \left[n_L\left(1-n_R\right)\Exp{\I\lambda}+n_R\left(1-n_L\right)\Exp{-\I\lambda}
      \right]\biggr\} \, ,
      \end{split}\end{gather}
where $n_{L,R}(\omega)$ are Fermi distribution functions in the left(right) lead with chemical potentials $\mu_{L,R}=\pm V/2$ and $T(\omega)$ is the effective transmission coefficient
\begin{gather}
      T(\omega)=\frac{\Gamma^2}{f(\omega)^{-2}+\Gamma^2}\text{.}
\end{gather}
with $\Gamma=2\pi\rho_0\gamma^2$. The function $f(\omega)$ has different expansions in the regimes $T=0$ and $T>0$.
\begin{gather}
f=\begin{cases} \Exp{-\alpha^2}\sum\limits_{n\geq 0} \frac{\alpha^{2n}}{n!} \frac{1}{\omega-\epsilon_0-n\Omega}& T=0, \\  \Exp{-\alpha^2\left(2n_\text{B}+1\right)}\hspace{-0.2cm}\sum\limits_{n=-\infty}^{\infty}\hspace{-0.2cm}\frac{I_n\left[2\alpha^2\sqrt{n_\text{B}\left(n_\text{B}+1\right)}\right]\Exp{n\Omega/2T}}{\omega-\epsilon_0-n\Omega} & T > 0.\end{cases}
\end{gather}
Again, in the case $T=0$, only processes involving the emission of phonons are allowed. In contrast, for $T > 0$, thermally excited phonons can be absorbed. The transmission coefficient $T(\omega)$ is made up of a sequence of peaks (see Fig.~\ref{fig:transcoeff}) at the energies $\mathbb{N}\Omega$ for $T=0$ or $\mathbb{Z}\Omega$ for $T>0$ and is properly normalized, $\int d\omega T(\omega)/\pi = 1$. By assuming a Lorentzian shape, the width of the peak at $\omega = n \Omega$ can roughly be estimated as $2\Gamma\Exp{-\alpha^2} \alpha^{2n}/n!$ and $2\Gamma\Exp{-\alpha^2(2n_\text{B}+1)} I_n[2\alpha^2\sqrt{n_\text{B}(n_\text{B}+1)}]$, for $T=0$ and $T> 0$ respectively.

Interestingly, both perfect and zero transmission are possible in the system. This is due to the special structure of $T(\omega)$, which is equivalent to the transmission coefficient of a system with an infinite number of spin-degenerate dots at energies $\epsilon_0 + n\Omega$ coupled to the leads in parallel (for the double dot system see, \eg, [\onlinecite{Kubala2003,Dahlhaus2010}]). This occurs as a consequence of the linear approximation: every single electron tunneling through the system takes along its polaron cloud, leaving the dot in exactly the same state as before the tunneling event. That means that the resonance condition is given by $\omega=\epsilon_0 + n\Omega$. The antiresonance (complete transmission suppression) emerges as an interference effect in precisely the same way as in the double-dot setup.\cite{Kubala2003,Dahlhaus2010} The above physical picture implies that the dwell time of electrons on the dot, which is on the order of $\Gamma^{-1}$, must be long compared to the inverse of the phonon (de)excitation rate. Since the latter is roughly proportional to $g$, we expect $g \gg \Gamma$ to be a necessary requirement for the validity of our approximation. This indeed implies a \emph{strong} electron-phonon coupling.

\begin{figure}[t]
\includegraphics[width=8cm]{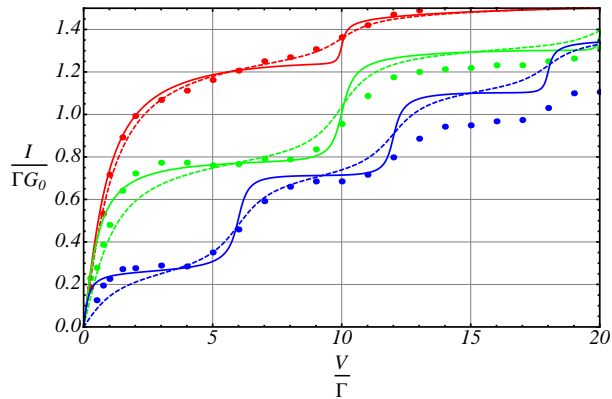}
\caption{(Color online) Current in single particle approximation, linear approximation and Monte-Carlo (MC) simulation data of Ref.~[\onlinecite{Muhlbacher2008}] for $T/\Gamma=0.2$. The dashed lines represent the single particle approximation as proposed in Ref.~[\onlinecite{Flensberg2003}], the solid line is the linear approximation and the dots represent the diagMC data. The energy level of the dot is held at $\epsilon_0=0$. The assignment of the data (from top to bottom) is the following $(g/\Gamma=2,\  \Omega/\Gamma=5,\ \text{red})$, $(g/\Gamma=4,\ \Omega/\Gamma=5,\ \text{green})$ and $(g/\Gamma=4,\ \Omega/\Gamma=3,\ \text{blue})$. The systematic deviation of the analytical results from the diagMC data for large $V/\Gamma$ is due to finite bandwidth necessary to carry out numerical simulations.}
\label{fig:current}
\end{figure}

At all temperatures one observes an exponential suppression of the peak width with the coupling $\alpha^2 = (g/\Omega)^2$. This leads to the well known Franck-Condon blockade, where sequential tunneling is exponentially suppressed and tunneling accompanied by phonon absorption/emission is preferred.\cite{koch05} This is also observable in the current. In Fig.~\ref{fig:current}, the current is depicted for different electron-phonon coupling constants $g$. For increasing coupling the step heights modifies nonuniformly and transport through states with higher $n-$phonon excitation, $n\approx (g/\Omega)^2$, gets more pronounced. The effect of temperature is similar to the phonon coupling strength $\alpha$ (this is obvious, because $\alpha^2\left(2n_\text{B}+1\right)$ or $2\alpha^2\sqrt{n_\text{B}\left(n_\text{B}+1\right)}$ always acts as an effective coupling strength). In order to assess the quality of our approximation we compared the calculated $I-V$ characteristic with the one from diagrammatic MC data of Ref.~[\onlinecite{Muhlbacher2008}]. In the regime of small to moderate $V$ and $\Omega$, as well as for $g>\Gamma$, our scheme indeed turns out to yield a better approximation than the single-particle approximation.\cite{Flensberg2003}

\begin{figure}[t]
\vspace*{0.3cm}
\includegraphics[width=8cm]{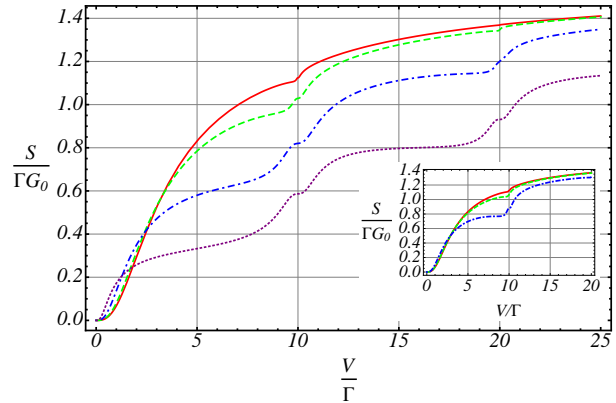}
\caption{(Color online) \emph{Main graph:} the noise in the linear approximation is depicted at $T=0$ for  $g/\Gamma = 1, 2,4,6$ (solid, dashed, dashed-dotted, dotted line, respectively). The remaining parameters are set to $\Omega/\Gamma=5$ and $\epsilon_0=0$. \emph{Inset:} the noise for temperatures $T/\Gamma=1,5,25$ (solid, dashed, dashed-dotted , respectively). The remaining parameters are $g/\Gamma = 1$, $\Omega/\Gamma=5$ and $\epsilon_0=0$.}
\label{fig:noise}
\end{figure}

The noise is plotted in Fig.~\ref{fig:noise}. Similar to the $I-V$ characteristic, is also shows a step-like behavior.
For strong coupling there is an additional feature: in the steps we observe an additional plateau. For large voltages the noise approaches the usual unitary limit, in our units $S = I$. 
Unsurprisingly, our approximation, being valid for not too large currents, shows no sign of enhanced noise due to the predicted avalanche-like transport behavior.\cite{koch06} To include higher-order correlations (see for example the \textit{nearest-neighbor crossing approximation} for the current in Ref.~[\onlinecite{zazunov2007}]) would be a task for the future.

In conclusion, we developed an approach to calculate the FCS of the Holstein polaron dot in a strong coupling regime. Using a linear approximation, we derived an analytical Levitov-Lesovik formula for the cumulant generating function with an effective, properly normalized transmission coefficient. Our approach yields predictions for zero temperature as well as for finite temperature, where the phonon is assumed to be thermally equilibrated.

The authors would like to thank H.~Soller, K.~F.~Albrecht and L.~M\"uhlbacher for many interesting
discussions and communicating the diagMC data. The financial support was provided by the DFG under
grant No.~KO~2235/3, by the Kompetenznetz ``Funktionelle
Nanostrukturen III'' of the Baden-W\"urttemberg Stiftung
(Germany), by the Swiss NSF, and by the HGSFP and CQD of the University of Heidelberg.

\end{document}